\documentstyle[12pt]{article}
\begin{document}
\begin{flushright}
SJSU/TP-93-11\\December 1993\end{flushright}
\vspace{1.7in}
\begin{center}\Large{\bf The Projection Postulate of Quantum Mechanics
 on the Lightcone}\\
\vspace{1cm}
\normalsize\ J. Finkelstein\footnote[1]{Participating Guest, Lawrence
        Berkeley Laboratory \\
       \hspace*{\parindent} \hspace*{1em}
        e-mail: JFINKEL@sjsuvm1.sjsu.edu}\\
        Department of Physics\\
        San Jos\'{e} State University\\San Jos\'{e}, CA 95192, U.S.A
\end{center}
\begin{abstract}
We discuss an interpretation of the projection postulate that implies
collapse of the wavefunction along the lightcone.
\end{abstract}
\newpage
In a recent article, Mosley [1] has proposed that the projection
postulate of quantum mechanics be interpreted to imply collapse
of the quantum-mechanical wave function along the past lightcone
of an observer, rather than along an equal-time hyperplane.
Mosley contrasts this proposal with others [2,3,4] in which the
collapse is taken to occur along a future light cone; he writes
``There have been some suggestions in the literature concerning
collapse along the forward lightcone...what might be described as
a progressive collapse.  We propose that collapse occurs immediately
over the entire past lightcone when a measurement occurs at any
point on it.''

This quotation does make it appear that there is a conflict between
the proposal made in [1] and that made in [2,3,4].  In this note
we will see that, in spite of this appearance, the
proposal made in [1] is completely compatible with (at least)
the proposal that the present author made in [4]; in fact, they can be
considered to be two alternative ways of describing the same
picture of wave-function collapse. We discuss in this note that picture which
emerges jointly from these two references.

The postulate that the wave function does collapse when a
``measurement'' is performed is, at best, highly controversial.
(See, for example, ref. [5].)
In this note, let us not enter into this controversy; we merely
wish to discuss the essentially kinematic question of how the same
picture of a collapsing wavefunction could be
alternatively characterized as occurring along the
past lightcone and along the future lightcone.  For this purpose,
we can naively accept the notion that a well-localized ``measurement
event'' can cause the wave function to collapse.

Let us begin with a very brief summary of the two proposals.
Mosley [1] uses a formalism [6] in which any given observer describes the
state of a system by a wavefunction $\psi (T)$ which is defined over
one of the past lightcones whose vertex lies on the world-line of
that observer.  Mosley then extends this formalism by proposing that
if a measurement event occurs somewhere along a particular past lightcone,
the wavefunction collapses along that entire past lightcone.
In ref. [4], it is observed that, if observers in different Lorentz frames
are to agree on the state of a system which is subject to the projection
postulate, then that state must depend on the position of the observer;
the notation $\psi_{\bf d}(t)$ was used to denote the state as described from
position ${\bf d}$ at time $t$. It was then suggested that a
measurement event would cause  $\psi_{\bf d}(t)$ to collapse when
${\bf d}$ was on the future lightcone of the measurement event.

Both ref. [1] and ref. [4] note that their proposed wavefunction
$\psi$ could be identified with the information about a quantum
system that could be available at a particular point. In fact, Malin [2]
has suggested that the wavefunction can {\em only}
be interpreted in terms of available information,
but whether or not that is true, the terminology
of information can be useful in clarifying the kinematics of the proposed
picture of collapse. Suppose we are concerned with the information
which is available at space-time point $O$ about events which take place at
space-time points $E_i$;
let us call $O$ the ``observation point'' and the $E_i$ the
``event points.''  If information were to travel at infinite speed,
there would be no dependence on the position of the
observation point; in the usual (non-relativistic) picture,
the position variables on which the wavefunction depends are
those of the event points. With information traveling
at the speed of light, information about a measurement event at $E$
is available at an observation point $O$ if $O$ is within the future
lightcone of $E$; this is the sense in which collapse (which we can
identify with updating of information) occurs along the future lightcone
of $E$.  But we can also say that, for a given observation point $O$,
the available information is that coming from events in the past lightcone
of $O$; this is the sense in which collapse of the wavefunction for a
given $O$ occurs along the past lightcone of $O$.

In the notation $\psi_{\bf d}$ used in [4], {\bf d}
represents the spatial components of the observation point. The notation
$\psi (T)$ used in [1] does not explicitly display the observation point.
Nevertheless, it is clear from the discussion in [1] that $\psi (T)$
{\em does} depend on the position of the observer; what we are here calling
the observation point is, within the formalism of [1], the vertex of the
past lightcone over which $\psi (T)$ is defined. Thus, we can identify
the picture of collapse suggested in [4] with that suggested in [1] by
identifying the ${\bf d}$ in  $\psi_{\bf d}$ as used in [4]
with the position of the vertex of the
lightcone discussed in [1].  This joint picture
can be described by the following two statements: first, that a measurement
event causes collapse of the wavefunction
at observation points which are on the future lightcone of the measurement
event, and second, that the wavefunction for a given
observation point (whether or not it collapses) is defined by events
within the past lightcone of that observation point.

Let us illustrate this picture with a simple example (also considered
in refs [1] and [4]): two electrons named $A$ and $B$, initially in an
entangled spin state; we take
\begin{equation}
\psi_{in} = {\textstyle \frac {1}{\sqrt{2}}}(|\uparrow _{A}\rangle
|\downarrow _{B}\rangle - |\downarrow _{A}\rangle | \uparrow _{B} \rangle).
\end{equation}
We suppose that the two electrons, as well as any observers we will consider,
are all at rest with respect to each other, so that we may use the
coordinates of their mutual rest system.  The positions of $A$ and $B$ are
respectively ${\bf r}_A$ and ${\bf r}_B$; define $\tau \equiv |{\bf r}_{A}
-{\bf r}_{B}|/c$. Now suppose that a measurement of the spin of $A$ at time
$t_A$ (and of course at position ${\bf r}_A$) produces the result
$\uparrow _A$; then the projection postulate implies that, after the
measurement, the state of $A$ and $B$ is no longer entangled as it was
in eq. (1), but rather becomes the product of pure states
$|\uparrow _{A}\rangle | \downarrow _{B} \rangle $.

Now let us pose some questions on which one might have expected refs.
[1] and [4] to differ: At what time does the state of $B$ become a pure state?
Is it at time $t_{A}+\tau$ (i.e., collapse along the future lightcone of
the measurement)? Or is it at time $t_{A}-\tau$ (i.e., collapse along the
past lightcone)?  In fact, in the picture presented in [1] and [4], these
questions are ambiguous; one cannot meaningfully speak of {\em the} state
of $B$, without specifying the observation point.  For an observation point
located very close to ${\bf r}_B$,
the collapse takes place at time $t_{A}+\tau$, since
the collapse occurs at observation points along the future lightcone of
the measurement event.  For an observation point located very close to
${\bf r}_A$, the collapse takes place at time $t_A$.  And if there
happens to be a clock located very close to ${\bf r}_B$,
then the observer at ${\bf r}_A$ simultaneously (at time $t_A$)
receives information
about the measurement and about the fact that that clock strikes the
time $t_{A}-\tau$. For an observation point located mid-way between
${\bf r}_A$ and ${\bf r}_B$, the collapse takes place at time $t_{A}+\tau /2$,
which is the same time as the arrival of information that the clock at
${\bf r}_B$ struck the time $t_A$.

Finally, let us re-state the picture of collapse that has emerged from
refs. [1] and [4]:  The wave function for a given observation point
corresponds to events within the past lightcone of that observation
point; a measurement event cause collapse for observation points
along the future lightcone of that measurement event.

\vspace{0.5cm}
Acknowledgement: I appreciate the hospitality of the
Lawrence Berkeley Laboratory, where this work was done.

\end{document}